# Room-temperature nanoseconds spin relaxation in WTe$_2$ and MoTe$_2$ thin films

*Qisheng Wang[1], Jie Li[2], Jean Besbas[1], Chuang-Han Hsu[3], Kaiming Cai[4], Li Yang[2], Shuai Cheng[2], Yang Wu[1], Wenfeng Zhang[2], Kaiyou Wang[4], Tay-Rong Chang[5], Hsin Lin[3], Haixin Chang[2]\* and Hyunsoo Yang[1]\**

[1]Department of Electrical and Computer Engineering, and NUSNNI, National University of Singapore, 117576 Singapore
[2]Center for Joining and Electronic Packaging, State Key Laboratory of Material Processing and Die & Mould Technology, School of Materials Science and Engineering, Huazhong University of Science and Technology, Wuhan 430074, China
[3]Centre for Advanced 2D Materials, National University of Singapore, 6 Science Drive 2, 117546, Singapore
[4]SKLSM, Institute of Semiconductors, Chinese Academy of Sciences, P. O. Box 912, Beijing 100083, China
[5]Department of Physics, National Tsing Hua University, Hsinchu 30013, Taiwan

**The Weyl semimetal WTe$_2$ and MoTe$_2$ show great potential in generating large spin currents since they possess topologically-protected spin-polarized states and can carry a very large current density. In addition, the intrinsic noncentrosymmetry of WTe$_2$ and MoTe$_2$ endows with a unique property of crystal symmetry-controlled spin-orbit torques. An important question to be answered for developing spintronic devices is how spins relax in WTe$_2$ and MoTe$_2$. Here, we report a room-temperature spin relaxation time of 1.2 ns (0.4 ns) in WTe$_2$ (MoTe$_2$) thin film using the time-resolved Kerr rotation (TRKR). Based on *ab initio* calculation, we identify a mechanism of long-lived spin polarization resulting from a large spin splitting around the bottom of the conduction band, low electron-hole recombination rate and suppression of backscattering required by time-reversal and lattice symmetry operation. In addition, we find the spin polarization is firmly pinned along the strong internal out-of-plane magnetic field induced by large spin splitting. Our work provides an insight into the physical origin of long-lived spin polarization in Weyl semimetals which could be useful to manipulate spins for a long time at room temperature.**

The emerging two-dimensional (2D) transition metal dichalcogenides (TMDs), due to their strong spin splitting in *d* orbits and valley momentum separation,[1] provide a fertile ground to explore the spin and valley degrees of freedom.[2, 3] The intrinsic inversion symmetry breaking in monolayer hexagonal 2H-TMDs ($MX_2$ with M=W, Mo and X=Se, S) generates the valley-contrasting optical selection rules. The spin-valley coupling reduces spin-flip scatterings, leading to a long spin and valley relaxation.[4] However, the semiconducting nature makes it challenging to apply a large current density in 2H-TMDs, restricting their applications in spintronic devices with typical vertical or lateral device structures of ferromagnet/TMDs.[5]

Recently, the semimetal 2D TMDs, e.g. Weyl semimetal $WTe_2$ and $MoTe_2$, have sparked intense research interest as they possess topological nontrivial electronic structures in the bulk and Fermi arcs at surface.[6] In stark contrast to 2H-TMDs, $WTe_2$ and $MoTe_2$ crystallize in the orthorhombic structure without centrosymmetry (1T′ phase).[7] With large spin-orbit coupling in W(Mo) 5*d* (4*d*) orbitals and Te 5*p* orbitals,[8] both the bulk and surface Fermi arcs present non-degenerate spin texture.[9] Interestingly, the intrinsic inversion-symmetry breaking of $WTe_2$ can induce an out-of-plane antidamping spin-orbit torque in $WTe_2$/ferromagnet heterostructures as the current is applied along the low-symmetry axis of $WTe_2$.[10] Meanwhile, the semimetallic $WTe_2$ and $MoTe_2$ have shown a remarkably high current density (~50 $MA/cm^2$),[11] indicating $WTe_2$ and $MoTe_2$ are promising candidates for generating large charge-to-spin current conversion. Therefore, understanding the underlying physical mechanisms of spin dynamics in $WTe_2$ and $MoTe_2$ is fundamentally important and technologically useful for semimetallic $WTe_2$ and $MoTe_2$ based spintronic devices.

Here, we investigate spin dynamics in centimeter-scale, chemical vapor deposition (CVD)-grown $WTe_2$ and $MoTe_2$ thin film by performing the time-resolved Kerr rotation (TRKR). We find a long lived spin polarization up to the nanosecond timescale at room temperature. A physical mechanism based on the first principle calculation of band structure is proposed to explain the long-lived spin polarization. Moreover, we find that the optically induced spin polarization is robust against an externally applied magnetic field, supported by spin dynamics analytical simulations.

WTe$_2$ and MoTe$_2$(space group $P_{mn21}$) have an orthorhombic structure consisting of 1D zigzag chains with W(Mo) atoms along the $x$ direction (Figure 1a). It possesses a layered structure like other two-dimensional (2D) TMDs, but with additional distortion of W(Mo) atomic chains along the $y$ direction (top panel in Figure 1a), resulting in an inversion symmetry breaking. The spin texture (Figure 1b, Figure S1, S2, S3a and b) and the spin-orbit splitting energy ($\Delta_{sp}$) (Figure 1c and Figure S3c) of two lowest conduction bands are estimated from the tight-binding Hamiltonian interpolated from the *ab initio* calculation (see Experimental Section for details). Due to the requirements of time-reversal symmetry and lattice symmetry operations ($c_{2z}$, $\sigma_{zx}$ and $\sigma_{yz}$), the spin polarization along $z$ ($S_z$) and $y$ ($S_y$) in the $k_x$ and -$k_x$ valleys have opposite signs (see Figure 1b, Figure S2a, Figure S3a, and S3b for $S_z$; Figure S1 and Figure S2b for $S_y$). For few-layers W(Mo)Te$_2$, the spin has no $x$-component because of the mirror symmetry with respect to the z-y plane ($\sigma_{zy}$). Moreover, few-layer WTe$_2$ and MoTe$_2$ share the same inversion asymmetric property as in bulk crystal, which leads to large spin-orbit splitting. The $\Delta_{sp}$ along the high symmetry path is defined by $E<S\uparrow>-E<S\downarrow>$, where the $E<S\uparrow>$ and $E<S\downarrow>$ indicate the energy bands with up and down spins, respectively. The $\Delta_{sp}$ at the bottom of the conduction band of bilayer and twelve-monolayer WTe$_2$ and MoTe$_2$ reaches 30-45 (Figure 1c) and 15-25 meV (Figure S3c) respectively, which is compareable to monolayer 2H-TMDs (Figure S4). The distinct spin orientation in few layers WTe$_2$ and MoTe$_2$ indicates the potential spin injection by circularly polarized light excitation.

The investigation of spin dynamics is performed on CVD-grown WTe$_2$ and MoTe$_2$ thin films. So far, most of WTe$_2$ and MoTe$_2$ thin film samples were mechanically exfoliated and thus limited to small flakes.[12, 13] Here we prepared centimeter-scale WTe$_2$ and MoTe$_2$ thin films via CVD (see Experimetal Section for details). The WTe$_2$ thin film is uniform in large area as shown in optical microscopy (OM) images (Figure 2a). The photograph reveals the thin film covers the whole substrate of Si/SiO$_2$ with the area of 1×1 cm$^2$ (inset of Figure 2a), which is important for practical applications and optical measurements. The WTe$_2$ thin film usually has ~5-11 monolayers with a typical thickness of ~6.4 nm measured by atomic force microscopy (AFM) (Figure 2b) for WCl$_6$-derived WTe$_2$ (sample 1). The Raman spectra (Figure 2c) show peaks at ~110, 116, 132, 163 and 211 cm$^{-1}$ corresponding to the $A_2^4$, $A_1^3$, $A_1^4$, $A_1^7$, and $A_1^9$ vibration modes, respectively, which agrees well with that of mechanically exfoliated thin flakes from WTe$_2$ single crystal (Figure S6) and previous reports,[14] demonstrating the 1T′-phase nature of WTe$_2$. The Raman mapping data of the $A_1^7$ characteristic peak at ~163 cm$^{-1}$ confirm the sample is in the uniform 1T′ crystal structure (inset of Figure 2c). The X-ray photoelectron spectroscopy (XPS) data in Figure 2d show the chemical states of W 4$d$ and Te 3$d$ electrons. The binding energy of W 4$d_{3/2}$ (256.8 eV), W 4$d_{5/2}$ (244.1 eV), Te 3$d_{3/2}$ (583.7 eV) and Te 3$d_{5/2}$ (573.4 eV) reflects the valence states of W (+4) and Te (-2) in WTe$_2$ thin film. X-ray diffraction (XRD) and high-resolution transmission electron microscope (HRTEM) results of WTe$_2$ are shown in Figure S9a and b, respectively. The XRD pattern of WTe$_2$ thin film is dominated by diffarction peaks of crystal planes (002$n$, n=1, 2, 3,......)[15], indicating the thin film orients along the z-axis (Fig. S9a). The HRTEM shows WTe$_2$ thin film is polycrystalline (Fig. S9b). It consists of multiple single crystallines with a grain size of dozens of nanometers. The lattice distance of (002) planes is ~ 0.7 nm.

A similar method (see Supporting Information Notes 1) was applied to prepare few-layer MoTe$_2$ (sample 2) with a thickness of ~11.4 nm (Figure S7b) using a molybdenum trioxide (MoO$_3$) thin film as a reaction source. The optical microscopy image (Figure S7a), Raman spectra (Figure S7c), Raman mapping (Figure S7d) and XPS (Figure S8) verify the pure and uniform 1T′ phase of MoTe$_2$ thin film.[12, 16] The quantitative analysis derived from the XPS spectra shows the atomic ratio of Te to W(Mo) in few-layer WTe$_2$ and MoTe$_2$ is ~2 which is consistent with the stoichiometric amount of WTe$_2$ and MoTe$_2$. The XRD (Fig. S9c) and TEM (Fig. S9d) results confirm the MoTe$_2$ thin film orients along the z-axis. The above two samples were used to perform TRKR.

The experimental TRKR setup is depicted in Figure 3a. Ultrafast laser pulses of a duration of 120 fs and wavelength of 800 nm were generated at a repetition rate of 1 kHz from a Ti: Sapphire oscillator. Each pulse was later split into an intense exciting pump pulse (800 nm) and detecting linearly polarized probe pulse (400 nm). All experiments were performed at room temperature (see Experimental Section). The pump and probe power are 600 and 20 µW, respectively. Figure 3b displays typical TRKR dynamics from sample 1 following the excitation by circularly polarized pump pulses. In the magneto-optical Kerr effect, the polarization of probe light is sensitive to the magnetization of the magnetic medium. In our work, the normal incident circularly polarized light (800 nm) induces a net out-of-plane spin polarization hold by the photo-generated carriers. The net spin polarization acts as a mean magnetization field which rotates the polarization of the probe light (400 nm) after reflection from the surface of WTe$_2$ and MoTe$_2$ thin films. In order to exclude the interference from the pump light (800 nm) on the Kerr signals, a long-pass filter was used to filter the pump light before the probe light (400 nm) enter the balanced photodetector (see Experimetal Section for the details).

Figure 3b shows that the Kerr signal changes its sign when the pump beam reverse its helicity ($\sigma^+$ and $\sigma^-$), emphasizing the fact that the light induced spin polarization is determined by the pump helicity. As shown in Figure 1b and Figure S1-S3, the semimetallic WTe$_2$ and MoTe$_2$ show distinct conduction and valence bands with $k$ dependent spin-splitting compared to that of metals due to the spin-orbit coupling. Here, we consider that the pump causes the transition of electrons from the occupied valence bands (W(Mo) $d$ orbitals) to the empty conduction bands (Te $p$ orbitals) and neglect the intraband transitions. We further consider that the angular momentum is conserved during the transitions. The spin dynamics of WTe$_2$ and MoTe$_2$ is reflected in the temporal decay of the Kerr rotation.

To avoid spurious experimental artifacts, we apply our analysis to the difference of the signals obtained with σ⁺ and σ⁻ pump pulses, as shown in Figure 3c. The resulting spin dynamics presents two decay processes, which can be estimated by a bi-exponential fitting. To elucidate the nature of long spin relaxation, we simplify the system to be governed by the spin-polarized electrons. Similar spin dynamics is observed in sample 2 (Figure S10). An initial fast exponential decay of $\tau_1$ = 33-61 ps ($t$< 0.1 ns) is followed by a slower decay of $\tau_2$ = 0.35-1.2 ns ($t$> 0.2 ns). The variation of lifetimes is likely caused by materials difference (0.35 ns for MoTe$_2$ and 1.2 ns for WTe$_2$). It is also challenging to keep the electronic properties such as carrier mobility and density of all thin films due to variable environmental factors in our CVD process. The spin relaxation time thus slightly changes from sample to sample.

Combining the spin texture of WTe$_2$ and MoTe$_2$ along Γ-X, we can understand the long spin relaxation as follows. The DFT simulations (Figure 1b, Figure S2a and S3a) have shown the spin polarization is almost out-of-plane where the pump light is applied. The time-reversal and lattice symmetry requires states with the opposite spins at $k_x$ and $-k_x$ valleys. Figure 3d is the schematic of WTe$_2$ band structure with spin relaxation process. The Hall transport data (Figure S11 and Supporting Information Notes 2) show the majority carrier of WTe$_2$ thin film is hole, indicating the Fermi level resides in the valance band. Electrons with up-spins are photogenerated by left-circularly polarized light (σ⁺) at $-k_x$ due to the conservation of angular momentum during the optical transition. Subsequently, a fast spin depolarization occurs with a characteristic time of 30–60 ps which originates from the carrier-phonon or carrier-carrier scattering in their respective bands at a rate $\tau_1^{-1}$. The recombination of electron-hole pairs decreases the spin density of photo-generated carriers causing the decay of Kerr effect signal. However, states at the conduction band and valance band are separated in momentum space (Figure 1b and Figure 3d). Therefore, the momentum conversion must be assisted by phonons and such a phonon-assisted process lowers the recombination

probability.Our results of ultrafast transient reflectivity suggest a weak phonon scattering in WTe$_2$ thin film. As shown in the inset of Figure 3c, subsequetly after light excitation, the photo-excited electrons undergo a fast carrier thermalization via carrier-carrier or carrier-phonon interaction. Then thermalized carriers (hot carriers)slowly relax with a decay time beyond our measurement range (2.7 ns). The long-lived hot carriers lifetime suggest a very weak electron-hole recombination rate in WTe$_2$ due to weak phonon scattering.In addition, WTe$_2$ possesses two valleys with opposite spin splitting due to time-reversal symmetry (Figure 1b and c). Thus it needs a momentum scattering from $k$ to $-k$ states for a spin depolarization. Such a process requires magnetic defects and is therefore profoundly reduced resulting in a long decay of the net spin polarization ($\tau_2 \sim 1.2$ ns).

Another interesting phenomenon observed in the spin dynamics of WTe$_2$ and MoTe$_2$ thin film is the robust spin lifetimes against an externally applied magnetic field ($B_{ext}$). The fluctuating spin-orbit coupling field $B_{So}$ alone will not dephase out-of-plane spins of electrons since $B_{So}$ orients along the $\hat{z}$-axis.[3] The external magnetic field $B_{ext}$ is applied at an angle of 60 degree with respect to the z-axis (inset of Figure 4b). Therefore, electron spins will precess about the total field, $(B_{SO}+B_{z(ext)})\hat{z} + B_{y(ext)}\hat{y}$,[17] where $B_z$ and $B_y$ are the components of the external magnetic field in the $\hat{z}$ and $\hat{y}$ direction, respectively. This is similar to momentum-dependent spin precession in conventional semiconductors such as the Dyakonov-Perel mechanism.[18] However, Figure 4a and S12 show that the Kerr signals of WTe$_2$ and MoTe$_2$ thin film are stable even when $B_{ext}$ increases to 618 mT. The spin lifetimes extracted from bi-exponential fitting in both WTe$_2$ and MoTe$_2$ thin film slightly vary with $B_{ext}$ as shown in Figure 4b. We propose that the spin stabilization in WTe$_2$ and MoTe$_2$ thin film originates from a large spin-orbit coupling field, which dominates over the external field. As evidenced by the theoretical calculation (Figure 1c and Figure S3c), the spin-orbit splitting energy ($\Delta_{sp}$) of two lowest conduction bands reaches 30-45 meV in bilayer and 15-25 meV in twelve-monolayers WTe$_2$ and MoTe$_2$, which respectively corresponds to an effective field $B_{so}$ of 260-389 and 130-216 T, dominating over $B_{ext}$. Analytical simulations (see Supporting Information Note 3 for the details) using a modified drift-diffusion model reveal that a large spin-orbit coupling field firmly stabilizes the spin polarization along out-of-plane $B_{so}$. As shown in Figure 4c and Figure S13, the spin precession occurs with a small $B_{so}$ ( $\leq 0.3$ T). However, the spin polarization oscillation disappears with a large $B_{so}$ ( $\geq 10$ T) and the effect of $B_{ext}$ on the spin relaxation is indeed negligible.

We compare the spin lifetimes in $WTe_2$ and $MoTe_2$ thin film with topological insulator $Bi_2Se_3$. The spin lifetimes of few-layer $WTe_2$ and $MoTe_2$ are three orders of magnitude longer than that of $Bi_2Se_3$ (< 3 ps)[19] at room temperature. The remarkable long-lived spin lifetimes in $WTe_2$ and $MoTe_2$ thin film is the consequence of unique band structures of $WTe_2$ and $MoTe_2$. The $WTe_2$ and $MoTe_2$ show the distinct spin splitting at states ($E$, $k$) and ($E$,-$k$) with opposite spin orientations. Topological protection from time-reversal symmetry suppresses the back momentum scattering from spin-up (down) ($E$, $k$) states to spin-down (up) ($E$,-$k$) states. In addition, the conduction band and valance band of $WTe_2$ and $MoTe_2$ are well separated in the momentum space, requiring phonons for the recombination of electrons and holes. Such a phonon-assisted recombination lowers the recombination probability of spin-polarized electrons and holes pairs. The low recombination probability of spin-polarized electrons and holes pairs is also confirmed by ultrafast reflectivity measurement which indicates a lifetime of photo-excited electron-hole pairs longer than 2.7 ns.

In conclusion, we observe the spin lifetime of nanoseconds at room-temperature in CVD-grown few-layer $WTe_2$ and $MoTe_2$ thin films using TRKR spectroscopy. We also find the spin polarization in $WTe_2$ and $MoTe_2$ thin films is robust against the external magnetic field. We further discuss the underlying physics of spin dynamics in $WTe_2$ and $MoTe_2$ thin film. Compared with topological insulators and 2H-TMDs, the semimetallic $WTe_2$ shows a remarkably high current density,[11] making $WTe_2$ and $MoTe_2$ promising for generating large charge-to-spin current conversion. In addition, the semimetallic nature of $WTe_2$ and $MoTe_2$ mitigates a current shunting issue through the ferromagnet in TMD/ferromagnet structures. Together with the long-lived and robust spin polarization, our results suggest WTe$_2$ and MoTe$_2$ may lead to novel spintronics applications.

**Experimental Section**

*Synthesis and characterization:* The few-layer WTe$_2$ and MoTe$_2$ thin film were synthesized in a three-zone CVD system (Figure S5). For WTe$_2$ thin film (sample 1) as a representative example, 0.3 g tungsten chloride (WCl$_6$) powders (99.99%, Alfar Aesar) and 0.4 g tellurium (Te) powder (99.99%, Alfar Aesar) were placed at the first and second zone, respectively. Silicon (Si) substrates with 300 nm silicon dioxide (SiO$_2$) were placed at the third zone. The furnace flowed by a mixture of 160 sccm N$_2$ and 40 sccm H$_2$ under ambient condition was heated to 500 °C for 20 mins. Synthesis details of sample 2 is provided in Supporting Information Notes 1. The CVD furnace was cooled down to room temperature naturally after the growth. To prevent the samples from oxidation, all thin films were protected by a vacuum-evaporated 1 nm Al capping which was oxidized to Al$_2$O$_3$ naturally in air. The samples were characterized by the HR800 Raman system (JY Horiba), AFM (SPM9700, Shimadzu Co.), XPS (AXIS-ULTRA DLD system with an Mg KαX-ray source), reflective OM (MV6100, Jiangnan Yongxin Co.), XRD (Empyrean, PANalytical B.V.) and TEM (Tecnai G2 F30, FEI Co.).

*Device fabrication and Hall transport:* The Hall devices were fabricated by two-step photolithography. In the first step, photolithography followed by the etching using Ar ion milling was performed to define the Hall bar. In the second step, the electrodes were defined by photolithography. Finally, the Ta (4 nm)/Cu (30 nm) contacts were sputter-deposited. The Hall transport was carried out by Quantum Design PPMS from room temperature to 2 K.

*Time-resolved Kerr rotation:* The TRKR measurements were carried out using a mode-locked Ti: Sapphire laser. The laser beam (800 nm) was split into a pump beam and a probe beam. The frequency of probe beam was doubled by second harmonic generation. The pump beam circular polarization was prepared by a succession of linear Glan-Taylor polarizer and quarter wave plate (QWP). The probe beam was linearly polarized by a Glan-Taylor polarizer. The pump beam was focused to a spot of ~300 μm through a lens and at normal incidence with respect to the sample surface. The probe beam (spot size ~100 μm) irradiated the sample at an angle of 10° from the normal direction of the sample surface. After reflection on the sample, the pump beam was filtered out by a long-pass filter from the reflected probe beam. The probe beam polarization was analyzed by a succession of half wave plate (HWP), Wollaston beam splitter (WBS) and balanced photodetector (BPD). The time delay between the pump and probe beam was adjusted with a mechanical stage. To allow lock-in measurements, the pump beam was modulated by a mechanical chopper with a frequency of 250 Hz. All the experiments were performed at room temperature. An external coil was used to apply magnetic field up to 683 mT.

*Ultrafast reflectivity:* The transient reflectivity were carried out using a mode-locked Ti:Sapphire laser. The femtosecond laser beam (800 nm) was split into a pump beam and a probe beam. Both the pump and probe pulses were focused onto the simple non-collinearly with spot diameters of about 300 and 100 μm, respectively. The differential reflectivity dynamics were recorded as a function of the time delay between the pump and probe pulses with cross-polarized.

*First principles calculations of spin texture:* The calculations were carried out by the VASP package based on the projector augmented wave method within the ultrasoft pseudopotential scheme. The generalized gradient approximation (GGA) is applied for the exchange-correlation interaction. In the calculations, spin-orbit coupling is considered self-consistently. From the first principles calculations, the Wannier tight-binding Hamiltonian of the bulk crystal is constructed based on the MLWF scheme. The thin film Hamiltonians with various layers can be built according to this interpolated tight-binding bulk Hamiltonian, and the band structures and spin expectation values were calculated via the standard way.

*Theoretical simulations of spin dynamics:* The simulations were derived from the drift-diffusion equation (Supporting Information Notes 3). The single-valley model was used based on the band structure of WTe$_2$ and MoTe$_2$ and assuming a *g*-factor of 2. The spin-relaxation rate within the band was set to 0.5-100 GHz. $B_{so}$ was divided into two distinct regions, conventional region ($B_{so} \ll B_{ext}$) and stabilization region by a large spin-orbit coupling field ($B_{so} \gg B_{ext}$).

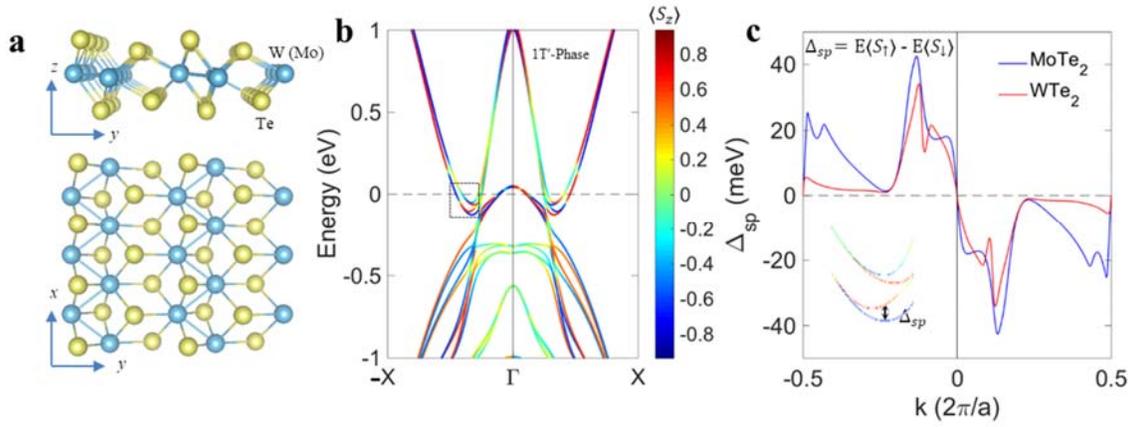

**Figure 1**. **Spin texture and spin-orbit splitting energy of WTe$_2$ and MoTe$_2$.** (a) Atomic structure of monolayer WTe$_2$ and MoTe$_2$. Light blue balls denote W(Mo) atoms. Yellow balls represent Te atoms. W(Mo) atomic chains distort along the $y$ direction (top panel). (b) Spin texture of bilayer WTe$_2$. The color bar in the right label indicates the out-of-plane spin polarization $<S_z>$. Due to the requirements of time-reversal symmetry and lattice symmetry ($\sigma_{zy}$), the spin reverses its polarizations at $-k_x$ and $k_x$. (c) The $\Delta_{sp}$ as the function of momentum $k_x$ along Γ-X. The $\Delta_{sp}$ of both WTe$_2$ and MoTe$_2$ at the bottom of the conduction band reaches ~40 meV. The inset shows the spin-orbit splitting energy $\Delta_{sp}$ on the conduction band because of inversion symmetry breaking. The $\Delta_{sp}$ is defined by energy band differences between spin-up (E$<S\uparrow>$) and spin-down (E$<S\downarrow>$) electrons. Only the results of bilayer are presented since $\Delta_{sp}$ is still obvious in both materials, when the thickness increases to twelve monolayers (Fig. S3).

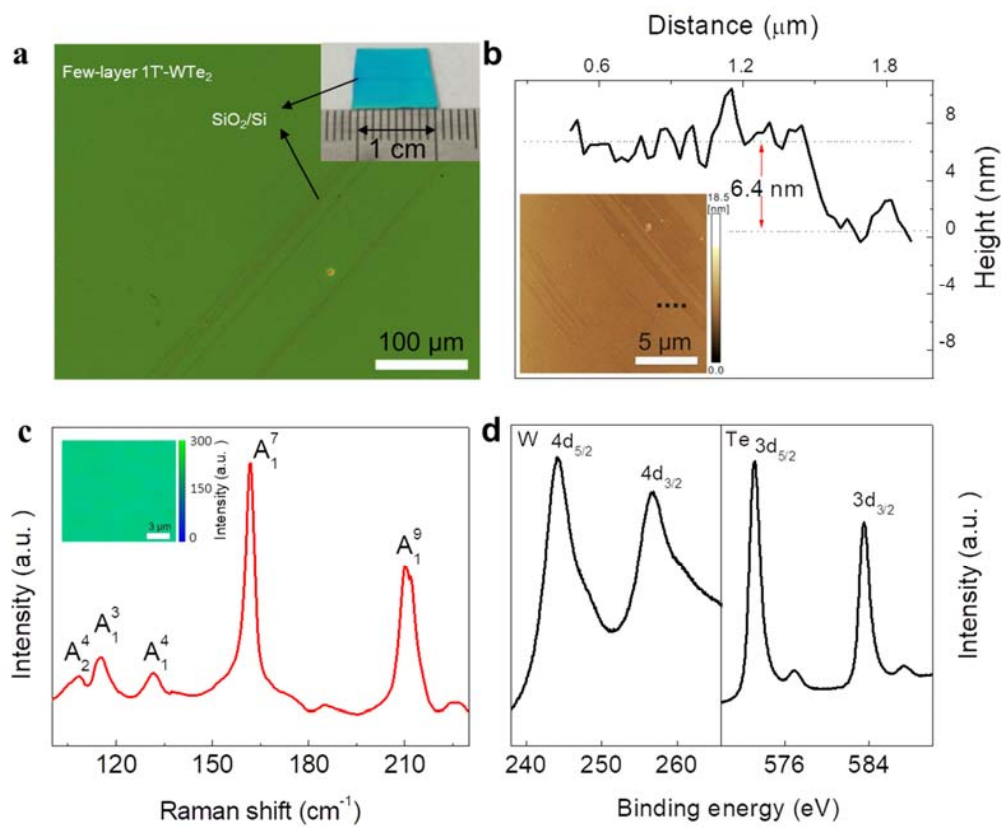

**Figure 2**. **Large-area CVD-grown WTe$_2$ thin film**. (a) OM image of a representative few-layer WTe$_2$ thin film. The inset shows a photograph of WTe$_2$ thin film on a Si/SiO$_2$ substrate. A scratch in the middle of photograph shows the contrast with Si/SiO$_2$ substrate. The WTe$_2$ thin film covers the whole Si/SiO$_2$ substrate with the area of 1×1 cm$^2$. (b) Height profile of WTe$_2$ thin film. The inset is the corresponding AFM image. The thickness ranges from 5 to 11 nm (~5-11 monolayers). (c) Raman spectra of a typical WTe$_2$ thin film. The peaks located at ~110, 116, 132, 163 and 211 cm$^{-1}$ corresponds to the $A_2^4$, $A_1^3$, $A_1^4$, $A_1^7$, and $A_1^9$ vibration modes, respectively, which indicates the 1T′-phase nature of WTe$_2$ thin film. The inset is the Raman mapping integrated from $A_1^7$ peak at 163 cm$^{-1}$, confirming the uniformity of WTe$_2$ thin film. (d) XPS from W 4$d$ and Te 3$d$ electrons. It shows the binding energy of W 4$d_{3/2}$ and W 4$d_{5/2}$ at ~256.8 and 244.1 eV, and Te 3$d_{3/2}$ and Te 3$d_{5/2}$ at ~583.7 and 573.4 eV, respectively. The atomic ratios of Te to W derived from XPS is ~2 which is consistent with the stoichiometry of WTe$_2$.

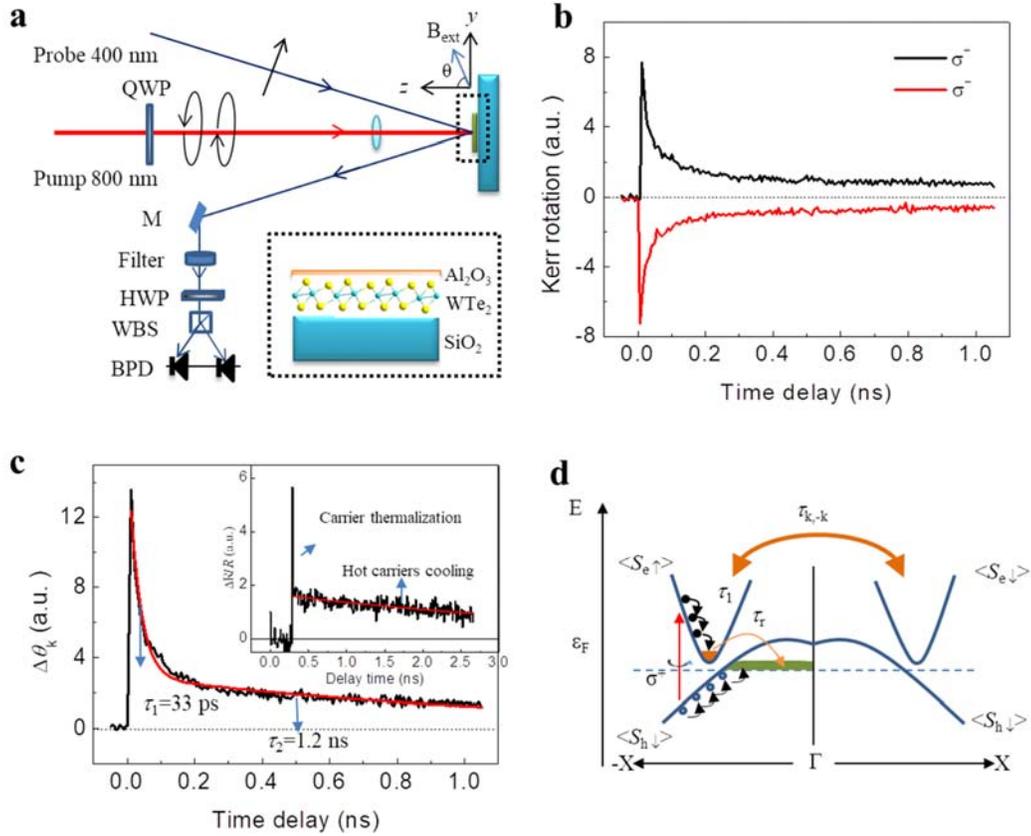

**Figure 3. Room-temperature long-lived spin polarization.** (a) Schematic diagram of TRKR setup. We use left- ($\sigma^+$) or right- ($\sigma^-$)circularly polarized pump pulses to excite spin-polarized electrons and holes. QWP: quarter wave plate. M: reflectivity mirror. HWP: half wave plate. WBS: Wollaston beam splitter. BPD: balanced photodetector. (b) TRKR traces under excitation of $\sigma^+$ and $\sigma^-$ pump. The Kerr rotation changes the sign when the helicity of pump pulse is reversed, indicating the Kerr rotation arises from optically induced spin polarization. (c) Signals difference ($\Delta\theta_k$) between $\sigma^+$ and $\sigma^-$ pump. Two dominant decay processes ($\tau_1$ = 33 ps and $\tau_2$ = 1.2 ns) can be extracted by bi-exponential fitting. Inset is the ultrafast transient reflectivity. (d) Schematic diagram of WTe$_2$ band structure with spin relaxation process. The momentum separation between the bottom of the conduction band and the top of the valence band obstructs the recombination of electron-hole pairs. Furthermore, the back-scattering between $k_x$ to $-k_x$ is forbidden due to time-reversal symmetry and lattice symmetry ($\sigma_{zx}$ and $c_{2z}$) operation. The horizontal dashed line shows the position of the Fermi level ($\varepsilon_F$). $<S_e\uparrow>$ and $<S_e\downarrow>$ denote spin-up and spin-down polarization of electrons, respectively, while $<S_h\uparrow>$ and $<S_h\downarrow>$ label spin-up and spin-down polarization of holes, respectively.

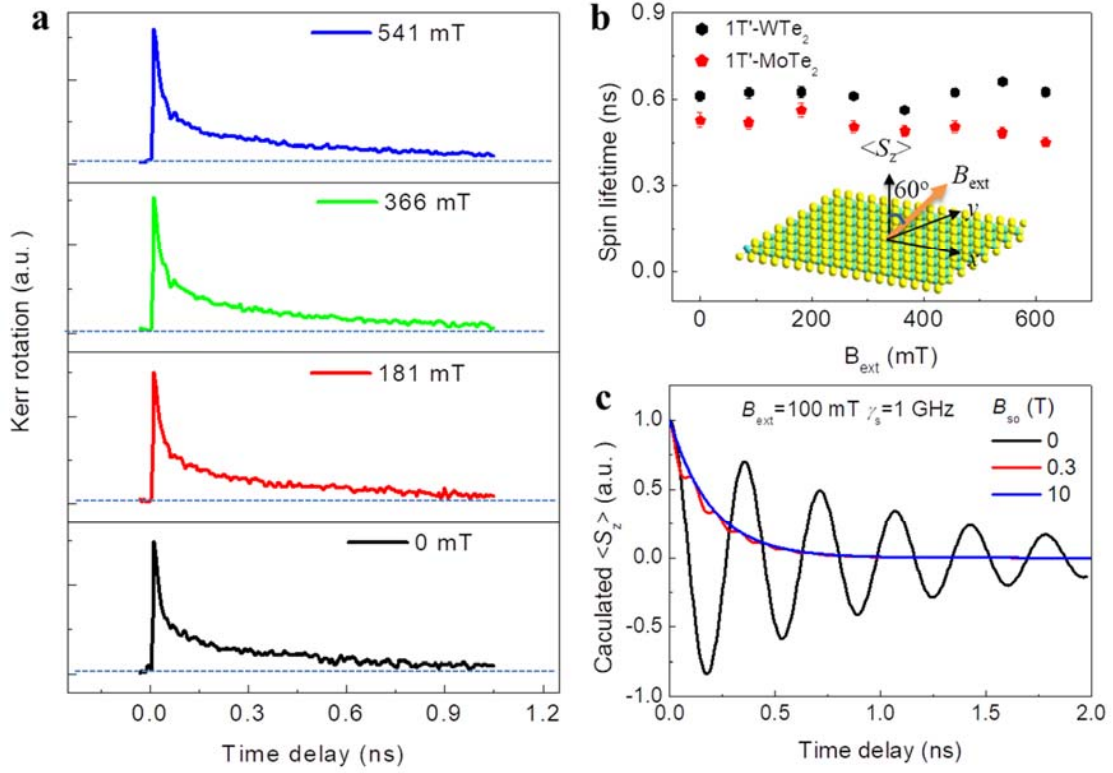

**Figure 4. Robust spin polarization against external magnetic field.** (a) Kerr rotation traces of few-layer WTe$_2$ thin film with $B_{ext}$. (b) The spin lifetimes extracted via bi-exponential fitting. The Kerr rotation almost remains unchanged as $B_{ext}$ increases, and the spin lifetime slightly varies with increasing $B_{ext}$. The inset is the schematic diagram showing the relative orientation of $B_{ext}$ and $S_z$. $B_{ext}$ is applied with an angle of 60° from the normal direction of the sample surface. (c) Simulation data of spin dynamics under different spin-orbit coupling fields. $\gamma_s$ represents the scattering rate at a given band. The obvious spin precession is observed under small $B_{so}$ ( ≤ 0.3 T). However, the spin polarization $<S_z>$ is stabilized when $B_{so}$ increases to 10 T (>>$B_{ext}$ = 100 mT). The large spin-orbit coupling fields (130-389 T) in WTe$_2$ and MoTe$_2$ thin films firmly pin optically induced spin polarization.